\RequirePackage{fix-cm} 

\documentclass[epjc3,english,pdftex,twocolumn]{svjour3}

\makeatletter
\def\makeheadbox{}
\makeatother
\date{}  

\usepackage{extarrows}
\usepackage{wasysym}
\usepackage{babel}
\usepackage{flushend}
\usepackage[T1]{fontenc}
\usepackage{graphicx}
\usepackage[switch]{lineno}
\usepackage{mathptmx} 
\usepackage[numbers,sort&compress]{natbib}
\usepackage{textcomp}
\usepackage{upgreek}
\usepackage{xcolor}
\usepackage{xspace}
\usepackage{booktabs}
\usepackage{multirow}
\usepackage{threeparttable}
\usepackage{colortbl}
\usepackage{arydshln}
\usepackage{makecell}
\usepackage{dcolumn}
\usepackage{csquotes}
\usepackage[normalem]{ulem}
\usepackage{comment}

\usepackage{amsmath}
\usepackage{amssymb}
\usepackage{amsfonts}

\usepackage[colorlinks,citecolor=blue,urlcolor=blue,linkcolor=blue]{hyperref}

\usepackage{microtype}

\journalname{Eur. Phys. J. C}
\smartqed  

\begin{document}

\title{\boldmath Application of surface coating for radon mitigation in rare-event searches}

\author{F. J\"org\thanksref{addr1, e1, e2}
        \and
        G. Eurin\thanksref{addr1, e3}
        \and
        H. Simgen\thanksref{addr1, e4}
        }

\thankstext{e1}{ now at Physik-Institut, Universit\"at Z\"urich, Winterthurerstrasse 190, 8057 Z\"urich, Switzerland }
\thankstext{e3}{ now at IRFU, CEA, Universit\'e Paris-Saclay, F-91191 Gif-sur-Yvette, France}
\thankstext{e2}{ e-mail: florian.joerg@physik.uzh.ch}
\thankstext{e4}{ e-mail: hardy.simgen@mpi-hd.mpg.de}

\institute{Max-Planck-Institut f\"ur Kernphysik, Saupfercheckweg 1, 69117 Heidelberg, Germany \label{addr1} 
}


\maketitle

\begin{abstract}
Rare-event searches offer a powerful avenue for investigating some of the most fundamental questions in modern physics, most prominently the particle nature of dark matter and the possible Majorana nature of the neutrino. Often, their dominant source of background comes from the radioactive noble gas radon emanating from materials.
We report on a novel strategy to mitigate this background by the application of coating layers. A method for electroplating of copper was developed that showed a thousandfold reduction of the $^{222}$Rn emanation rate from a $^{226}$Ra-implanted stainless steel sample.

\keywords{Radon barriers \and Coating \and Low radioactivity techniques \and Rare-event searches}

\end{abstract}

\section{Introduction}
\label{sec:intro}

\begin{figure*}
    \centering
    \includegraphics[width=0.75\textwidth]{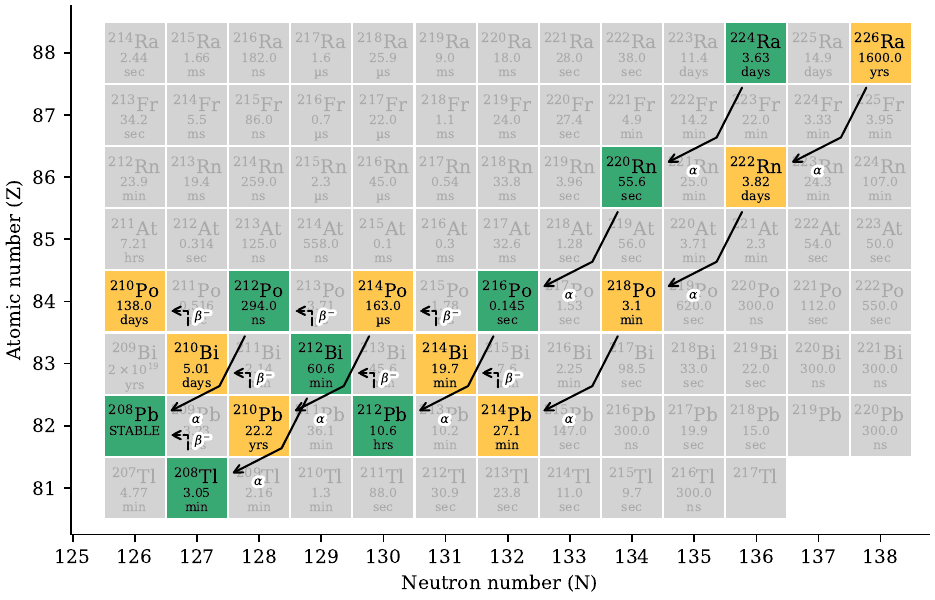}
    \caption{Excerpt of the primordial $^{238}$U (orange) and $^{232}$Th (green) decay chains, containing the radioactive noble gases $^{222}$Rn and $^{220}$Rn respectively. Only the parts relevant to this work are highlighted. All numerical values are taken from\,\cite{ENSDF}.}\label{fig:decay_chains}
\end{figure*}

There is hardly a place to be found, where the naturally occurring, radioactive noble gas $^{222}$Rn is absent. In our daily life, it is responsible for roughly half of the natural radiation exposure\,\cite{UNSCEAR:2018}.  $^{222}$Rn is so abundant, because it is  part of the primordial $^{238}$U decay chain (orange in \autoref{fig:decay_chains}) and traces of uranium are present in any material.  It is constantly produced from the alpha decay of $^{226}$Ra and may be released into the surrounding medium.
Therefore, small amounts of this isotope can also be found in detectors searching for extremely rare events such as the neutrinoless double beta decay\,\cite{Pas:2015eia} or the interaction of dark matter particles\,\cite{MarrodanUndagoitia:2015veg}. The decay of $^{222}$Rn is followed by a chain of radioactive daughter decays inducing background events, which can be confused with the rare signal events searched for. Similar arguments hold for the other naturally occurring isotopes $^{220}$Rn and $^{219}$Rn, but $^{222}$Rn is often the most dangerous isotope due to its relatively long half-life of 3.8 days.

Therefore, great care has to be taken for its mitigation. This is usually done by a careful pre-selection of the building materials of the detectors\,\cite{BOREXINO:2001bob, MANESCHG2008448, LEONARD2008490, LZ:2020fty, XENON:2021mrg, PandaX-4T:2021lbm}. Additionally, methods to actively remove radon from the detection media via adsorption\,\cite{Nakano2020} and distillation\,\cite{Bruenner2017, Aprile2017b, Murra:2022mlr} are applied and software tools are being developed to tag and remove $^{222}$Rn-induced events \cite{Rn_veto2024, LZ:2024zvo}.

This work was done in the framework of the series of XENON dark matter experiments. In XENONnT\,\cite{XENON:2024wpa} an unprecedented low $^{222}$Rn concentration of less than 1\,$\upmu$Bq/kg in liquid xenon\,\cite{XENON:2025nic} could be achieved. Despite this achievement, the demanding requirements for future xenon detectors such as the XLZD\,\cite{XLZD:2024nsu} (0.1\,$\upmu$Bq/kg) experiment are still challenging. Therefore, the existing radon suppression techniques must be complemented by novel radon mitigation methods.

\begin{figure}[t]
    \centering
    \includegraphics[width=0.4\textwidth]{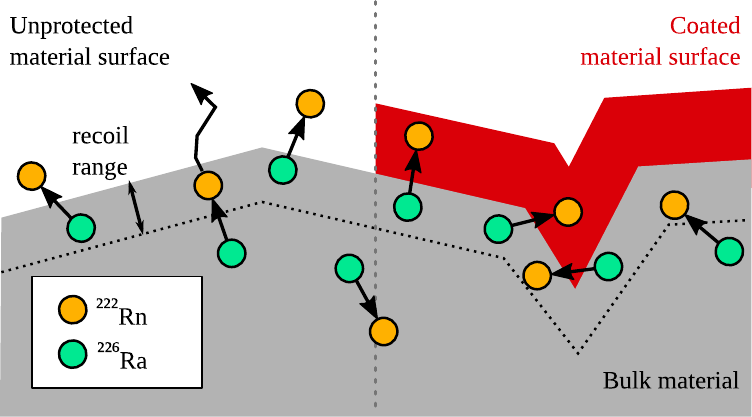}
    \caption{${}^{222}$Rn is released from detector materials surfaces via recoil and/or diffusion (left). The protective coating layer (right, red), seals the surface and prevents the release of radon into the surrounding medium.}\label{fig:coating_principle}
\end{figure}
In this work, we investigated the capability of different surface coating methods to form a sealing layer that reduces the radon release rate from materials as sketched in \autoref{fig:coating_principle}. Such a coating layer needs to fulfill several requirements:
(1) Its thickness must exceed the typical recoil range of the $^{222}$Rn daughter nucleus during its radioactive production process.
(2) The layer needs to be sufficiently tight against the diffusion of radon.
(3) It must have an extremely low $^{226}$Ra concentration by itself.
(4) A sufficient mechanical stability and adhesion is required even at liquid xenon temperature.
(5) The coating layer must have a low out-gassing behavior to preserve the high chemical purity, which is often required in particle physics detectors.

The measurement technique and method to evaluate different coating layers is described in \autoref{sec:measurement_technique}, followed by a description of the used samples in \autoref{sec:samples}.
A summary of the investigated coating techniques and the methods used to assess the quality of the coating layers is provided in \autoref{sec:coating_method} and \autoref{sec:analytics}.
The resulting radon reduction achieved by the different coatings are reported in \autoref{sec:results}, while \autoref{sec:conclusion} will give a summary and an outlook to the remaining challenges to be addressed before its application in a real detector.

\section{Measurement methods}
\label{sec:measurement_technique}

The performance of the investigated coating methods is estimated via the radon reduction factor $R$. It is obtained from the sample's radon emanation rate $A$ before and after the coating has been applied
\begin{align}
    R = \frac{A_\text{before}}{A_\text{after}}~.\label{eq:reduction_factor}
\end{align}

Reduction factors for the two radon isotopes $^{220}$Rn and $^{222}$Rn are evaluated in this study. Depending on the isotope, either low-background miniaturized proportional counters were used (for $^{222}$Rn)\,\cite{Zuzel2009,XENON:2020fbs} or an electrostatic radon monitor (for $^{220}$Rn)\,\cite{Brunner:2017xsu} were applied. The latter one can also determine $^{222}$Rn, but the proportional counters have better sensitivity.

For the measurements using the miniaturized proportional counters, the sample was stored for several days under an initially $^{222}$Rn-free helium atmosphere in order to allow for a sufficient accumulation of the emanated radon.
Afterwards, the accumulated radon is separated from helium using a liquid nitrogen cooled radon trap. Potential impurities such as oxygen or water are removed by a hot getter. Finally, the radon sample is filled into a miniaturized proportional counter along with an argon/methane (P10) gas mixture to determine the number of radon atoms from their decay. Due to the low intrinsic background and their high detection efficiency, these counters can detect very low $^{222}$Rn-activities as low as 20\,$\upmu$Bq\,\cite{XENON:2020fbs}.

The electrostatic radon monitor consists of a stainless steel vessel instrumented with a windowless Hamamatsu S3204-09 Si-PIN diode\,\cite{hamamatsu:2021}.
Following the alpha decay, the radon progeny has a high probability to be produced in a positively charged state\,\cite{Pagelkopf2003}, allowing for their electrostatic collection by a negative high-voltage potential of 1\,kV applied to the Si-PIN diode.
Subsequent alpha decays of these collected radon daughters can then be detected and counted in order to infer the rate at which radon is released from the sample. In our case the $^{220}$Rn activity is measured through its daughter nuclide ${}^{212}$Po, which has a higher detection efficiency compared to ${}^{216}$Po, the other alpha-decaying ${}^{220}$Rn daughter nuclide. More details on the setup and measurement method can be found in\,\cite{Jorg:2022spz}.

Besides its radon emanation rate, the total activity of a sample can be determined using gamma-ray spectrometry. For these measurements the high purity germanium (HPGe) spectrometer described in\,\cite{Budjas2008} was used, which is located at the shallow underground laboratory at the Max-Planck-Institut f\"ur Kernphysik\,\cite{Laubenstein:2004}.

\section{Sample substrates}
\label{sec:samples}

Surface coatings can be applied to many different types of materials. Despite the wide range of possible substrates, we have restricted our studies to stainless steel and thoriated tungsten. Thoriated tungsten welding electrodes (WTh, 4\,wt\% $\mathrm{ThO_2}$) are commercially available and emanate a significant amount of $^{220}$Rn (see decay chain in\,\autoref{fig:decay_chains}). It turned out that they even feature a small $^{222}$Rn emanation rate, stemming from trace impurities of $^{226}$Ra. Therefore, we developed our coatings on these electrodes.

\begin{figure}[h]
    \centering
    \includegraphics[width=0.49\textwidth]{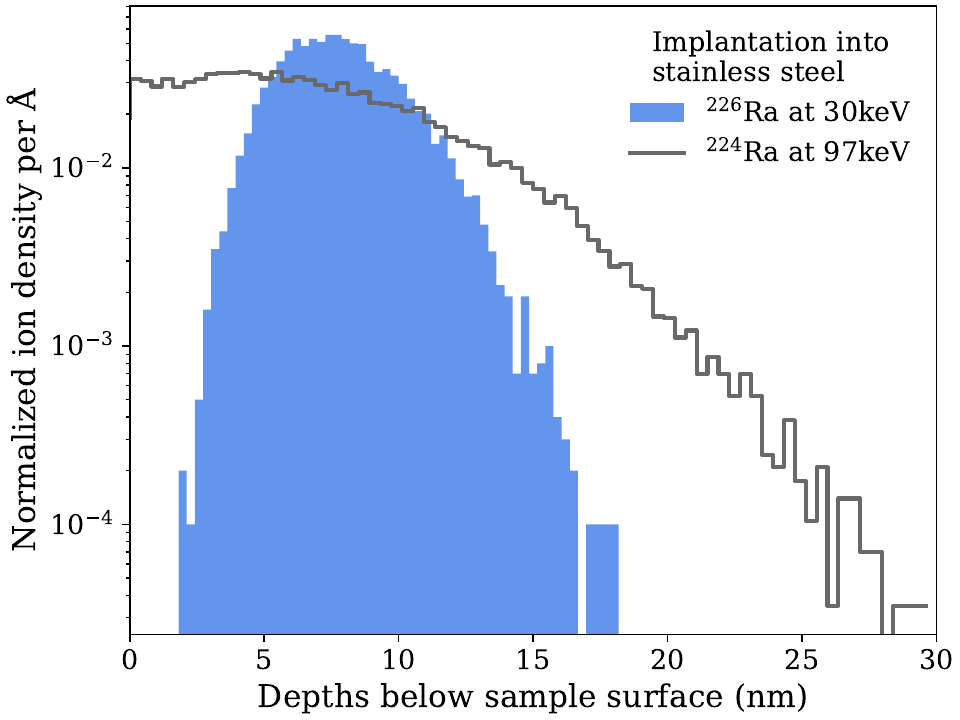}
    \caption{Simulated distribution of ${}^{224}$Ra and ${}^{226}$Ra ions after implantation using recoils from an extended ${}^{228}$Th source and a radioactive ion beam, respectively. The simulation is performed using SRIM\,\cite{Ziegler2010} and redrawn from\,\cite{Jorg:2022spz,Jorg:2022tli}.\label{fig:implantation_profile}}%
\end{figure}

Stainless steel substrates are more interesting, because stainless steel is widely used as detector construction material. However, the typical $^{222}$Rn emanation rate of stainless steel is too low to test the performance of coating layers. To overcome this problem, we custom-developed radon-emanating stainless steel samples. This was done by implanting them with radium at a depth of a few nanometers (see\,\autoref{fig:implantation_profile}), allowing a significant fraction of the produced radon to escape via the recoil received during the radium alpha decay. This method ensures that most of the activity is protected by a thin layer of substrate material, which is an advantage over other types of radon sources (e.g. electroplated ones).
Two different methods were used for this implantation:

$^{224}$Ra was implanted using the 97\,keV recoil energy from the alpha decay of $^{228}$Th.
For this, samples were placed next to an open 5\,kBq $^{228}$Th source\,\cite{Lang:2016zde} in a vacuum vessel.
Exposition of the stainless steel sample to the source for approximately one week under sub millibar pressure results in an implanted $^{224}$Ra activity of roughly 100\,Bq. 
Since $^{224}$Ra has a rather short half-life of 3.6\,days, the implanted activity is not permanent.
Nevertheless, the technique offers a simple way for a repeated production of short-lived $^{220}$Rn-emanating samples.
Because the recoiling $^{224}$Ra nuclei are emitted isotropic by the source, they can arrive at the stainless steel surface under shallow angles. Additionally, they loose part of their energy while traversing the active ${}^{228}$Th layer of the source. Both effects result in a reduction in the maximum implantation depth. The implantation profile was simulated using SRIM\,\cite{Ziegler2010} and is shown in gray in \autoref{fig:implantation_profile}. The median implantation depth was found to be 6.5\,nm\,\cite{Jorg:2022spz}.

The second type of sample is produced by implantation of $^{226}$Ra using the radioactive ion beam provided by the ISOLDE\,\cite{Kugler1992} facility located at CERN.
For this, $^{226}$Ra was produced by proton bombardment of an uranium carbide target. Next, the $^{226}$Ra was thermally released from the target, ionized and accelerated to an energy of 30\,keV and then shot into the stainless steel substrate, resulting in an average implantation depth of 7.9\,nm. The simulated profile is illustrated by the blue histogram in \autoref{fig:implantation_profile}.
Two sources, with an implanted activity of roughly 9\,Bq offering a $^{222}$Rn emanation rate of approximately 2\,Bq, were available for the work presented here. A more detailed description and characterization of these sources can be found in\,\cite{Jorg:2022tli,Jorg:2022spz}. Further such sources were produced recently\,\cite{Joerg:2023cds,Noia2024} allowing for more dedicated studies in the future.

\autoref{tab:sample_emanations} summarizes the different sample types and lists their typical radon emanation rates.

\begin{table*}[]
    \centering
    \begin{tabular}{cccc}
        \toprule
        Sample                    &   Material       &   $^{220}$Rn emanation &   $^{222}$Rn emanation\\
        \midrule
        Welding electrode ($\diameter$ 1\,mm) &   96$\%$\,Tungsten, 4$\%$\,Thorium        &   30\,mBq     &   0.04\,mBq\\
        Welding electrode ($\diameter$ 4.8\,mm) &   96$\%$\,Tungsten, 4$\%$\,Thorium      &   70 -- 90\,mBq     &   0.07 -- 0.14\,mBq \\
        $^{224}$Ra implanted plate (2cm$\times$2cm$\times$0.1cm)  &   Stainless steel 1.4435  &   40\,Bq     &   ---          \\
        $^{226}$Ra implanted plate (2cm$\times$2cm$\times$0.1cm)  & Stainless steel 1.4435  &   ---      &   2\,Bq       \\
        \bottomrule
    \end{tabular}
    \caption{Approximate equilibrium radon emanation activity for the different samples applied in this study.}
    \label{tab:sample_emanations}
\end{table*}

\section{Coating methods}
\label{sec:coating_method}

\begin{figure}[h]
    \centering
    \includegraphics[width=0.45\textwidth]{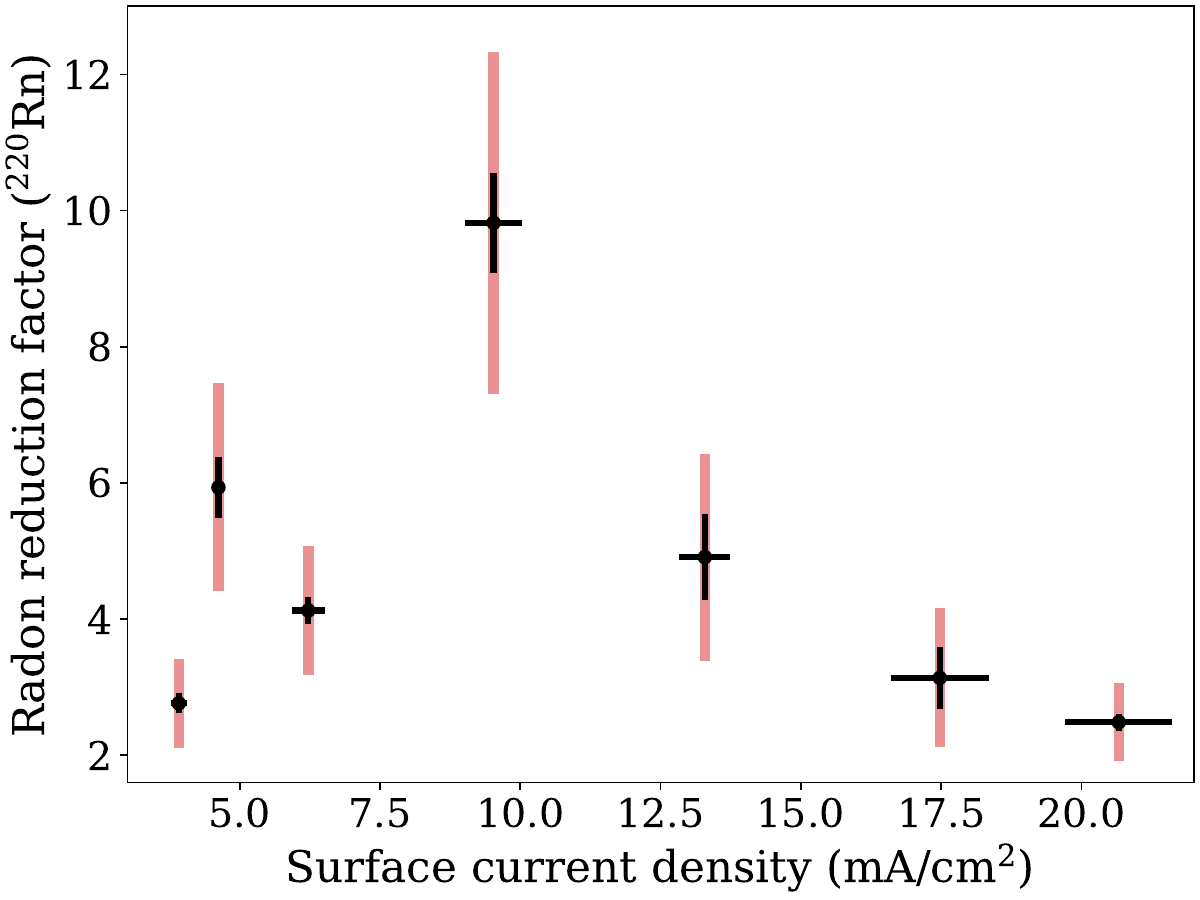}
    \caption{$^{220}$Rn reduction factors observed for coated thoriated tungsten electrodes as a function of the surface current density. The coating layers included in the Figure meet the quality criteria discussed in\,\autoref{sec:analytics}, and feature an optimum reduction at around 10 mA/cm$^2$\,\cite{Jorg:2022spz,Piotter2019}. The systematic uncertainty (light red) reflects the variation of initial activity between individual electrodes.}\label{fig:optimum_current}
\end{figure}

Various surface coating techniques exist for a wide range of applications, but sealing against inert gases is not a typical field of industrial application. Therefore, no prior guess could be made as to which coating processes promise the greatest chance of success for the particular case of radon suppression. We concentrated our efforts on in-house developed electrochemical plating of copper motivated by the extremely high purity that can be achieved for electro-formed copper\,\cite{Hoppe:2008xcd,Hoppe:2014nva,Suriano2018}. We produced layers with a thickness of approximately 5\,$\upmu$m from an electrolyte solution containing 0.05\,mol/l copper sulfate (CuSO$_4$) dissolved in 1\,mol/l sulfuric acid (H$_2$SO$_4$)\,\cite{WANG200372}. The electrolyte bath was heated to about $45^\circ$C and agitated using a stirrer. We optimized our procedure on thoriated welding electrodes. Among other parameters, we varied the surface current density and found that the highest radon reduction factor was obtained for 10\,mA/cm$^2$\,\cite{Piotter2019}. This dependence is shown in\,\autoref{fig:optimum_current}. Once satisfied with the coating procedure, we adapted the same recipe to stainless steel substrates. Here, the adhesion was weaker and we had to slightly modify the procedure. First, we created a 1\,$\upmu$m thick adhesion layer by applying a higher surface current density of 50\,mA/cm$^2$. The other 4\,$\upmu$m were added on top of the adhesion layer using the standard recipe. The hybrid coating layer showed sufficient adhesion to stainless steel and was used for all stainless steel samples presented in this work.

Besides electro-plating, we also investigated the following promising alternative techniques in cooperation with industrial partners:
\begin{itemize}
    \item Titanium coating by physical vapor deposition / magneton sputtering (PVD/MS)\,\cite{Swann:1988} in cooperation with \textit{EC Europ Coating GmbH (Germany)}\,\cite{Europcoating}.
    \item Copper coating by vacuum plasma spraying (VPS)\,\cite{Niessen:2011} in cooperation with \textit{Dr. Laure Plasmatechnologie GmbH (Germany)}\,\cite{Laure}.
    \item Amorphous hydrogenated carbon coating by chemical vapor deposition (CVD)\,\cite{Franceschini:2002} in cooperation with \textit{Innovative Coating Solutions SA (Belgium)}\,\cite{InovativeCoatingSlutions}.
\end{itemize}

For all the tested techniques, we were able to achieve mechanically stable coating layers on thoriated tungsten welding electrodes. However, the measured radon reduction factors were not competitive. We therefore did not pursue these methods any further. We note that the quality of the vapor-deposited layers might be improved by coating with other elements or by the application of multi-layer coatings in which layers of different elements alternate or by higher process temperatures, which are expected to create more compact coating layers than in our case\,\cite{Thornton:1977}. More details on the investigation of those methods can be found in\,\cite{Fischer:2016, Joerg2017, Lecher2019, Piotter2019}.

\section{Assessment of layer quality}
\label{sec:analytics}

After the development of a recipe for thoriated tungsten and stainless steel, we assessed the quality of the copper layer. First, they had to pass a simple visual inspection for optical appearance and homogeneity. Second, an adhesion test was done using a piece of sticky tape, pressed to the coated substrate and rapidly pulled off. For adhesive coatings, there must not be any residuals of the layer sticking to the tape. All samples coated using our optimized procedure passed the adhesion tests. To further stress the samples and to simulate effects of thermal cycles, we shock-cooled the coated samples by immersing them in a bath of liquid nitrogen and them warmed them up to room temperature. The cycle was repeated ten times. Again, an adhesion test was applied afterwards and no change with respect to prior the thermal cycles was observable.

\begin{figure}[h]
    \centering
    \includegraphics[width=0.49\textwidth]{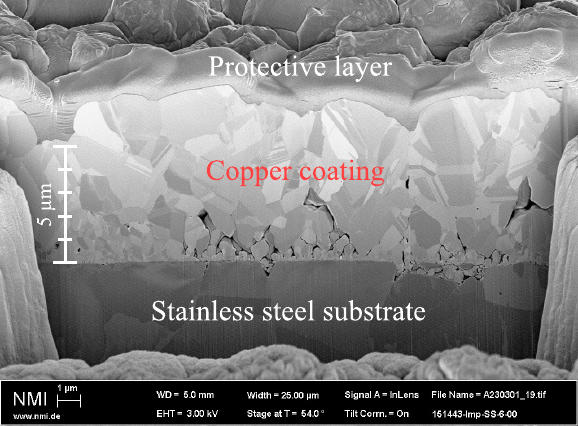}
    \caption{Scanning electron microscopic (SEM) image of a cross-section through an electroplated copper layer deposited onto stainless steel. The cross section was prepared using a focused ion beam (FIB). The thin platinum protective layer was added to prevent image artifacts induced by the ion beam.}\label{fig:sem_image}
\end{figure}

\autoref{fig:sem_image} shows a cross-sectional view of our coating layer.
The cross section was prepared via focused ion beam milling (FIB), and imaged using a scanning electron microscope (SEM) by the Naturwissenschaftliches und Medizinisches Institut an der Universit\"at T\"ubingen (NMI). 
A thin platinum protective layer was deposited \textit{in-situ} via electron- and ion beam bombardment of an injected precursor gas. This was done to prevent image artifacts (curtaining) typically caused by the ion beam milling.
The homogeneous looking stainless steel substrate is visible in the lower part of the picture, with the deposited copper layer above. 
The thickness of the copper layer matches the expected thickness of roughly 5\,$\upmu$m and consists of individual grains with sizes ranging from some hundreds of nano meters to several micro meters.
Close to the stainless steel surface, the average copper grain size is smaller and the layer features more voids between the grains than further away from the stainless steel substrate. 
This is likely caused by the about 1\,$\upmu$m thick adhesion layer, deposited at a higher current density. 
While that layer provides good adhesion of the coating to the stainless steel, it appears to be insufficient for blocking radon diffusion due to the many voids. 
At further distance from the surface, however, the coating becomes more dense and forms a closed layer with improved quality.
It should be stressed that the image was taken at an arbitrary selected spot of the sample and it is not guaranteed that it is representative for the whole layer.

Furthermore, the $^{228}$Th in the thoriated tungsten electrodes allows for an \textit{in-situ} validation of the recoil tightness of the coating layer. Following its alpha decay, the $^{224}$Ra progeny nucleus receives a recoil energy of 97 keV, which can result in its ejection from an uncoated sample, or from a coated sample suffering from an imperfect coating layer. The liberated $^{224}$Ra is then implanted into the alpha detector, where it can be identified by its characteristic alpha decay energy. As expected, the $^{224}$Ra signal was clearly visible in uncoated samples, whereas no signal was visible for coated thoriated samples. This holds for both, electrochemically plated and vapor-deposited coatings and suggests that the coatings are able to suppress recoil-driven release of nuclei. Consequently, the fraction of radon emanation, which is driven by recoil only, will also be suppressed.

\section{Results and interpretation}
\label{sec:results}

As explained in\,\autoref{sec:measurement_technique} we can give suppression results for the two radon isotopes of interest.
The upper half of \autoref{tab:results_methods} (sample No. 1 -- 3) shows the results for thoriated welding electrodes. The achieved reduction factors for $^{220}$Rn exceed a factor of 100. There is also an indication of a non-vanishing $^{222}$Rn reduction factor.
\renewcommand{\arraystretch}{1.2}
\begin{table*}
    \centering
    \caption{Summary of the achieved results for electrochemical plating (ECP).}
    \label{tab:results_methods}
    \begin{threeparttable}
        \begin{tabular}{lc|ccc}
            \toprule
            &	& \multicolumn{2}{c}{Reduction factor}    &   \\
            \multirow{-2}{*}{No.} &  \multirow{-2}{*}{Substrate}	&\multirow{1}{*}{$^{222}$Rn}	&	\multirow{1}{*}{$^{220}$Rn}	    &   \multirow{-2}{*}{Source}\\
            \midrule
            1  &   one thoriated tungsten electrode & -     &   107$\pm$14   &  \cite{Joerg2017} \\
            2  &   one thoriated tungsten electrode & -     &   47$\pm$4.3   &  \cite{Joerg2017} \\
            3  &  seven thoriated tungsten electrodes & $7.4^{+2.5}_{-1.5}$     &   129$\pm$3   &   \cite{Joerg2017}\\
            \hline
            4  &   $^{224}$Ra-implanted SS  & -   &   20 - 84   &  \cite{Jorg:2022spz} \\   
            5  &   $^{224}$Ra-implanted SS  & -   &   143-640   &  \cite{Jorg:2022spz} \\   
            6  &   $^{224}$Ra-implanted SS  & -   &   12.3-1050   &  \cite{Jorg:2022spz}\\ 
            7  &   $^{226}$Ra-implanted SS  &   $1500 \pm 70\,\mathrm{(stat)} ^{+250}_{-50} \,\mathrm{(syst)}$\,\tnote{1} & -  &  \cite{Jorg:2022spz}\\
            \bottomrule
        \end{tabular}
        \begin{tablenotes}
            \item [1] Value after equilibrium is reached (see \autoref{fig:isolde_coated})
        \end{tablenotes}
    \end{threeparttable}
\end{table*}	
\renewcommand{\arraystretch}{1}

\begin{figure}[h]
    \centering
    \includegraphics[width=0.49\textwidth]{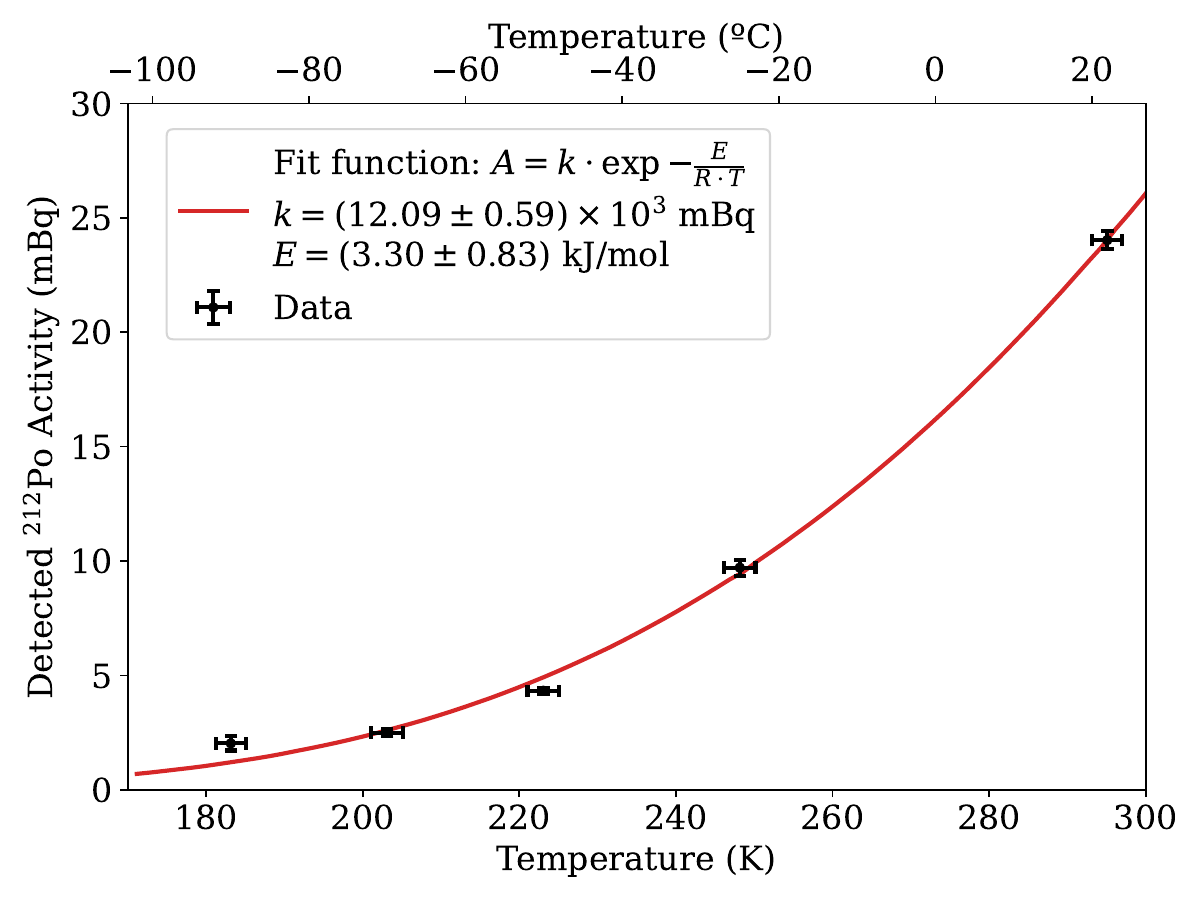}
    \caption{Low temperature dependence of the $^{220}$Rn emanation rate from thoriated tungsten electrodes coated with an electrodeposited copper layer. Data redrawn from \cite{Lecher2019}.} \label{fig:low_temp}
\end{figure}

\begin{figure}[h]
    \centering
    \includegraphics[width=0.49\textwidth]{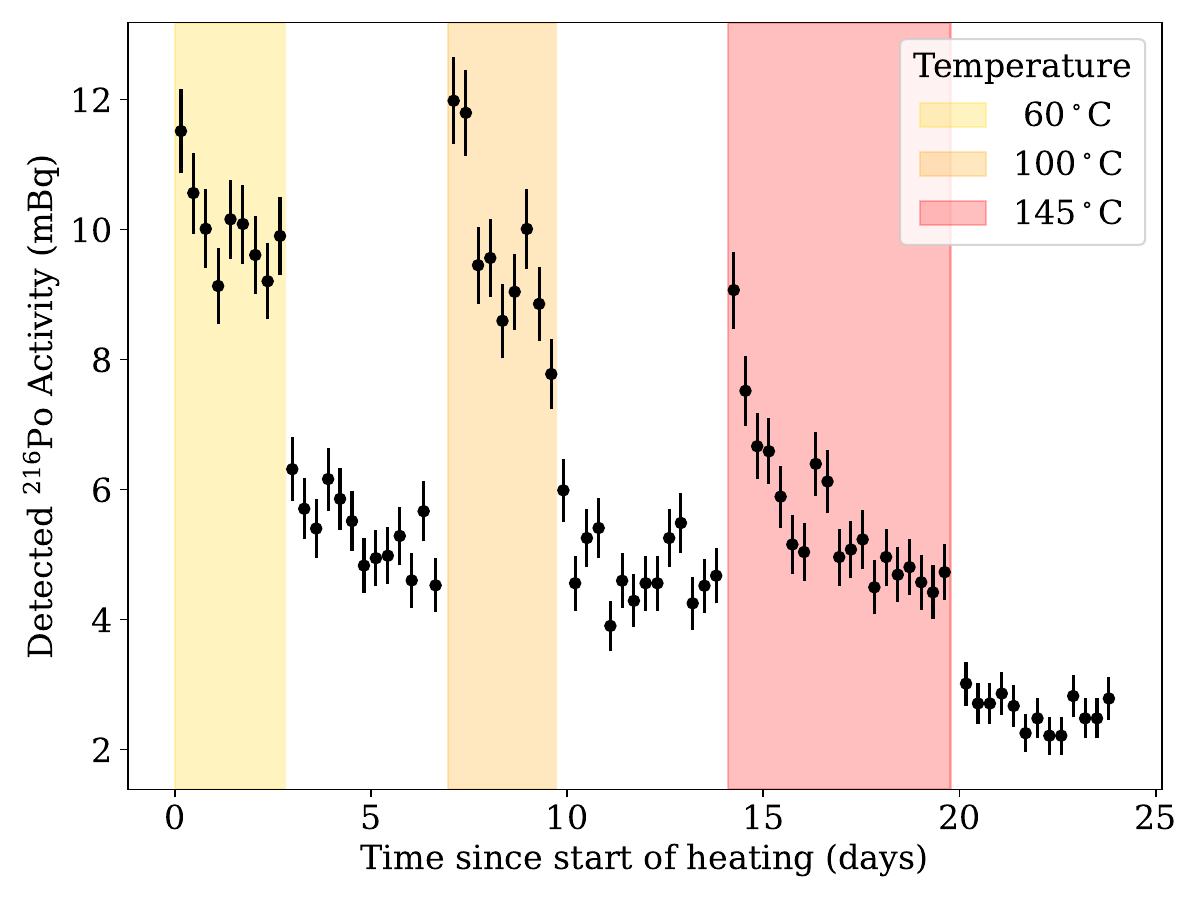}
    \caption{Effect of elevated temperatures onto the release of $^{220}$Rn from electroplated copper coatings deposited onto thoriated tungsten electrodes. The heat treatment results in a persistent reduction of emanation rate attributed to annealing of the copper. Data redrawn from\,\cite{Piotter2019}.}\label{fig:temp_evolution}
\end{figure}

We also studied the thermal dependency of the coating layer tightness. For that purpose, we placed ten of the coated electrodes in a vessel which can be cooled down to -90 $^\circ$C and heated up to 150 $^\circ$C. The sample vessel was connected to a radon monitor. Nitrogen was used as a carrier gas and an integrated pump guaranteed continuous recirculation of the carrier gas. The $^{220}$Rn emanation rate of coated thoriated tungsten electrodes decreases if the temperature is lowered (see\,\autoref{fig:low_temp}). The $^{220}$Rn emanation rate follows the Arrhenius law \cite{DiffusionSolidBook}, which is characteristic for diffusion-driven emanation. The results for elevated temperature are shown in\,\autoref{fig:temp_evolution}. Colored bands indicate regions, in which the temperature was raised. As expected from the Arrhenius law, the $^{220}$Rn emanation rate increases when the temperature is enhanced. But the rate is not stable during phases of elevated temperature. Instead the activity drops during the heating phase and the decrease is stronger if the temperature is higher. We interpret this observation as an annealing of the coating layer. The freshly created coating layer may be subject to mechanical tension. Under thermal treatment, the tension is released by re-arrangement of copper atoms in the layer. This will eventually lead to a more densely packed copper layer, which is a tighter barrier against radon diffusion. Also, the $^{220}$Rn emanation rate after the end of the thermal treatment is lower than before. This is another indication for a permanent re-structuring of the coating layer and confirms that annealing takes place.

\subsection{Radium-implanted stainless steel samples}

Samples No. 4 -- 8 in\,\autoref{tab:results_methods} show the results obtained for the radon reduction of coated stainless steel samples. Note that the coating procedure slightly differs from the procedure for thoriated tungsten. As described in\,\autoref{sec:coating_method} an adhesion layer had to be added to improve the stickiness of the copper layer on stainless steel substrates. With the modified procedure, the adhesion was good and all samples passed the tape test (see\,\autoref{sec:analytics}).

The $^{220}$Rn source distribution of the $^{224}$Ra-implanted stainless steel samples is in a substantial way different from the thoriated welding electrodes. In the latter one the $^{228}$Th and thus the $^{224}$Ra is homogeneously distributed, whereas the recoil-implanted $^{224}$Ra is at very shallow depth (see\,\autoref{fig:implantation_profile}) and it cannot be excluded that part of the activity is lost in the electrolyte during the coating process. A subsequent observation of a $^{220}$Rn emanation rate reduction could be due to the good performance of the coating in terms of radon retention. It could also be due to the loss of $^{220}$Rn sources into the electrolyte. Therefore, only an upper limit on the radon reduction factor due to the coating can be derived.
To constrain the radon reduction factor from the other side, we remove the coating after the measurement of radon reduction. The de-coating is done with the same setup by reversing the electrical voltage. In case there was no loss of $^{224}$Ra, the original $^{220}$Rn emanation rate will be restored after the de-coating. Any deviation is attributed to a loss of $^{224}$Ra into the electrolyte and a lower limit can be derived from the comparison of initial $^{220}$Rn emanation rate and the emanation rate after de-coating. 

\begin{figure}[h]
    \centering
    \includegraphics[width=0.49\textwidth]{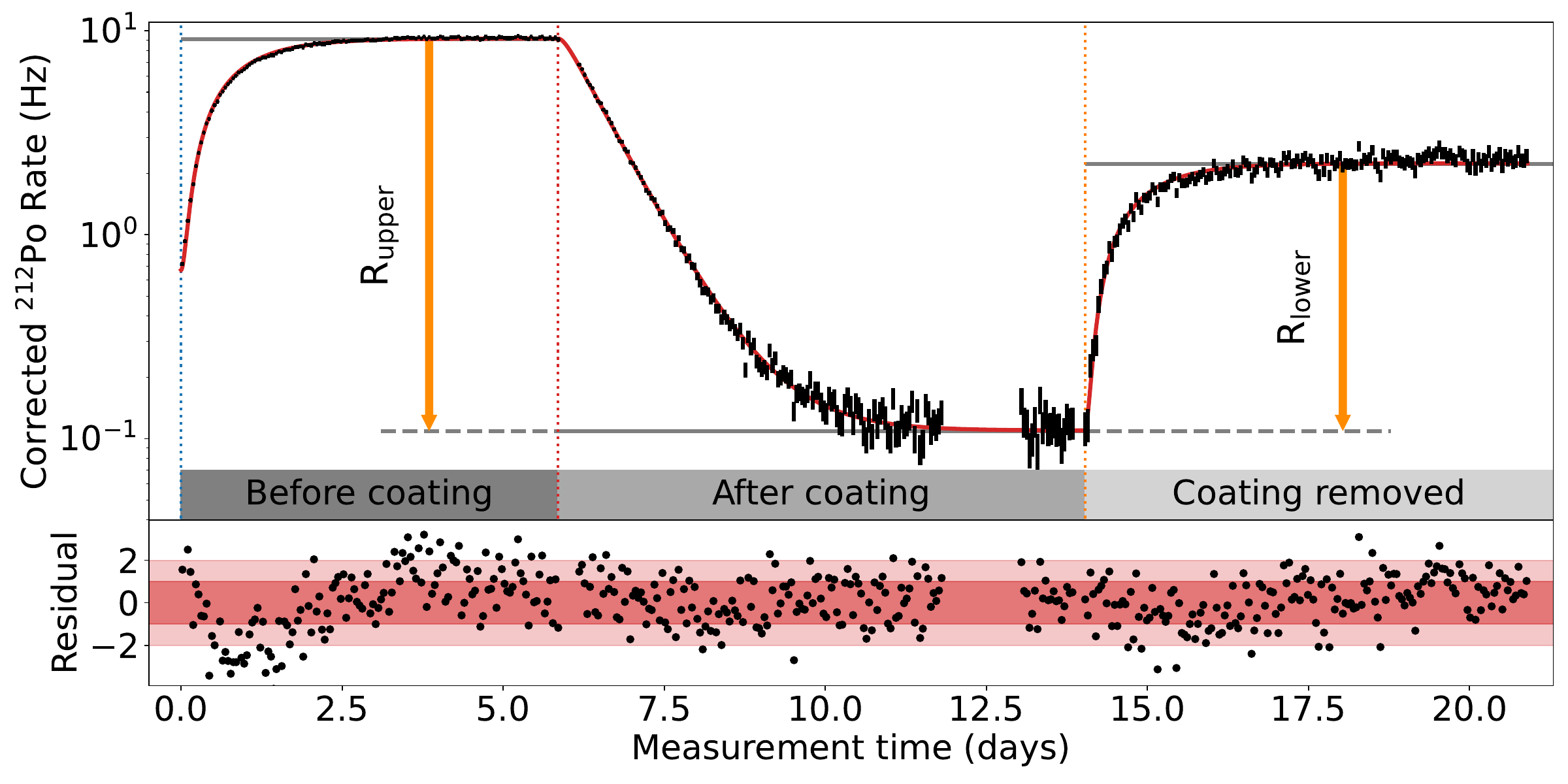}
    \caption{Time evolution of the detected $^{212}$Po alpha activity from a recoil-implanted $^{224}$Ra sample prior to application of the coating layer (left), with the coating layer applied (center) and after removal of the coating layer (right). For clarity, the data has been corrected for the loss of the mother isotope $^{224}$Ra due to its decay. The rate evolution of $^{212}$Po follows the expectation from the radioactive decay chain (red line).}\label{fig:implanted_cycle}
\end{figure}

\autoref{fig:implanted_cycle} shows a full measurement cycle for a $^{224}$Ra-implan\-ted stainless steel sample. $^{224}$Ra is short-lived, but its radioactive decay can easily be calculated. The data in \autoref{fig:implanted_cycle} is corrected for the loss of activity due to that decay. First, the initial $^{220}$Rn emanation rate is determined through the measurement of the $^{212}$Po decay rate (see\,\autoref{sec:measurement_technique}). After a couple of days equilibrium is reached. Then, the measurement is interrupted and the sample is coated. Afterwards, the rate reduction due to the coating can be observed until a new lower equilibrium is reached. Finally, the coating is removed and the counting rate recovers. It doesn't reach its initial value due to the partial loss of $^{224}$Ra activity from the sample mentioned above. Such a loss into the electrolyte was observed for all coated $^{224}$Ra-implanted samples. It could be confirmed by a HPGe spectrometry measurement of the used electrolyte.

The entire data set is fitted\,\cite{batemansolution} to determine the lower and upper limit for the radon reduction factor due to the coating. The quality of the fit is good except for the first few days, where a slight systematic deviation is noted. This is attributed to detector effects related to the radon daughter collection efficiency. 
The measured range of $^{220}$Rn reduction factors for three $^{224}$Ra-implanted samples is given in\,\autoref{tab:results_methods}. In all cases, the $^{220}$Rn reduction factor is equal or better than for thoriated tungsten.

\begin{figure}[h]
    \centering
    \includegraphics[width=0.49\textwidth]{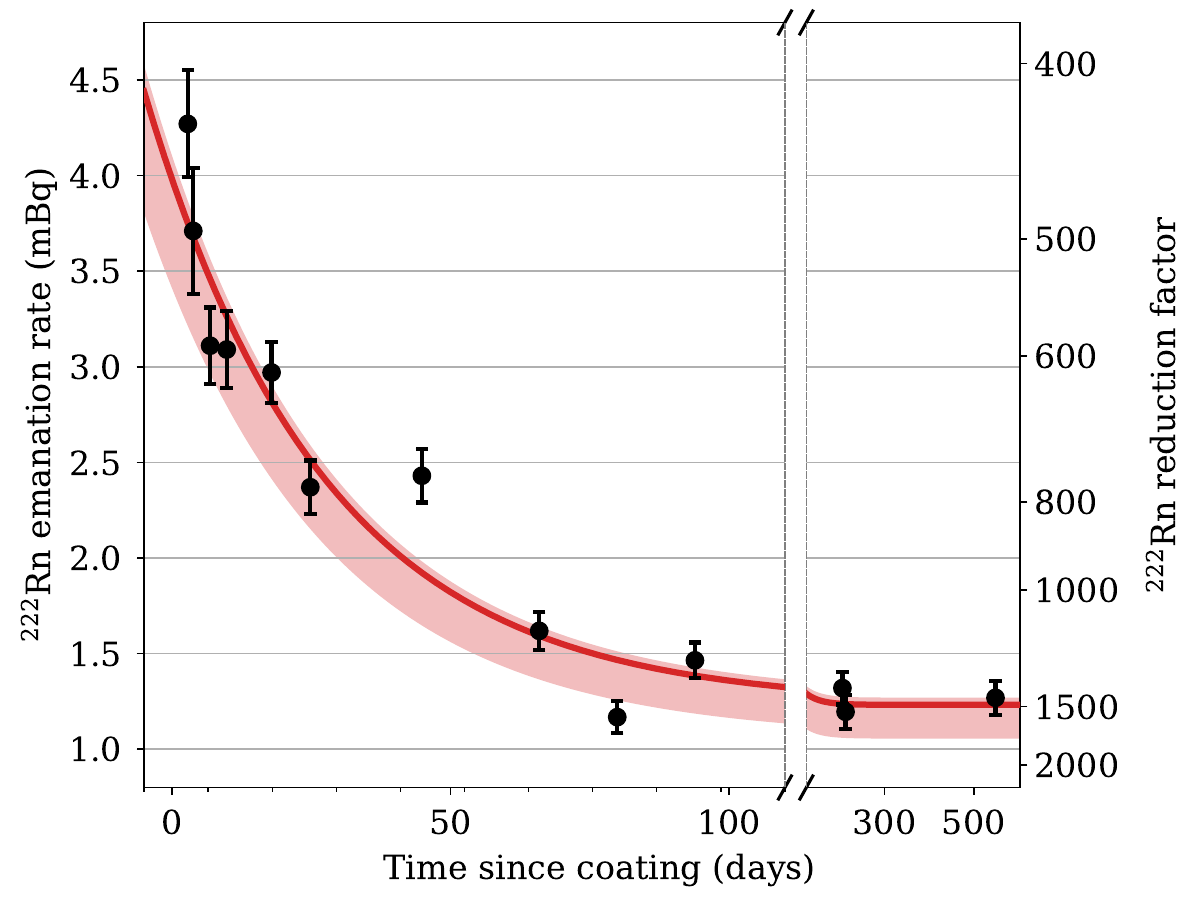}
    \caption{$^{222}$Rn emanation rate of sample 7 ($^{226}$Ra implanted stainless steel) as a function of time after the application of a 5\,$\upmu$m thick copper coating. The decrease is well described by an exponential function (red) and attributed to room temperature self annealing of the copper. Error bars indicate statistical uncertainty, while systematic uncertainty is indicated by the red shaded region.}\label{fig:isolde_coated}
\end{figure}

The final sample, which was coated, is a $^{226}$Ra-implanted stainless steel sample (No. 8). The sample represents the most interesting case, as it simulates the emanation of the most dangerous isotope $^{222}$Rn. It was coated using the very same recipe as used for the $^{224}$Ra-implanted samples. 
The $^{222}$Rn emanation activity of the $^{226}$Ra-implanted sample was measured before and after the coating with proportional counters. The reduction factor was initially found to be 470 and was monitored for almost 2 years afterwards. \autoref{fig:isolde_coated} presents the temporal development of the activity, which shows a decreasing trend, which reached a stable value after about half a year. This effect is attributed to the self-annealing of the coating layer. Although there was no thermal treatment, as in the case of thoriated welding electrodes, grain growth in electrodeposited copper layers has been reported to occur even room temperature\,\cite{self_annealing}. The increase in grain size results in a reduction of grain boundaries and internal stress, which leads to an enhanced diffusion tightness.
The time constant of this process is determined from a fit of the data with an exponential function and found to be $(32\pm4)$\,days
The final $^{222}$Rn-reduction factor for the coated $^{226}$Ra-implanted stainless steel sample is $1500 \pm 70\,\mathrm{(stat)} ^{+250}_{-50}\,\mathrm{(syst)}$ (see\,\autoref{tab:results_methods}), which is the highest observed radon reduction factor of all the coatings presented in this work.

Similar to the case of the $^{224}$Ra implanted samples, possible removal of implanted radium during the coating process adds a source of uncertainty. The comparison of HPGe measurements performed before and after coating revealed a reduction of the implanted activity by about 8\%. The reduction factor is corrected for this loss, while a possible variation in the distribution of activity after the coating within the sample is reflected in the systematic uncertainty (see \cite{Jorg:2022spz} for further details). In general, the loss of $^{226}$Ra activity was found to be much lower compared to the $^{224}$Ra-implanted samples. This is explained by the fact that the distribution of $^{226}$Ra does not extend all the way to the surface of the sample, as is the case for $^{224}$Ra (see \autoref{fig:implantation_profile}). Despite the lower implantation energy of only 30\,keV (97~keV for the case of $^{224}$Ra), this is a consequence of the $^{226}$Ra beam being perpendicular to the stainless steel substrate, whereas it is isotropic in the case of $^{224}$Ra implantation.

\section{Summary and outlook}
\label{sec:conclusion}

The sensitivity of astroparticle physics experiments searching for rare events has been steadily increased over the past years. This progress was possible because the experiments not only used an ever-increasing target mass, but at the same time the radioactive background was continuously reduced by sophisticated suppression methods. In liquid xenon detectors, external background became subdominant and gave an increased relevance to the xenon intrinsic radio impurities. Most dominant among those is $^{222}$Rn released from material surfaces, making radon mitigation a very important concept for the next generation of LXe detectors searching for direct dark matter interactions as well as for the neutrinoless double beta decay.

We have investigated surface coatings as radon barriers as a novel approach to address this problem. Among the various investigated coating techniques, electrochemical plating of copper showed the best performance. The approximately 5\,$\upmu$m thick coating layer completely suppressed recoil-driven $^{222}$Rn emanation and reduced diffusion-driven $^{222}$Rn emanation significantly. Overall, a $^{222}$Rn reduction factor of more than three orders of magnitude was achieved, which already far exceeds the requirements for future experiments.

While the obtained results demonstrate the feasibility of surface coating as a novel radon mitigation technique, this is only the first step towards the application of this technique in future experiments. Several challenges, such as demonstration of the scalability of the developed procedure to larger surfaces as well as demonstration that the demanding constraints regarding chemical- and radio purity of the coatings will be addressed in future studies. For the latter, it is hoped to incorporate the impressive radiopurity results obtained with electroformed copper that were demonstrated e.g. by\,\cite{Hoppe:2008xcd,Hoppe:2014nva,Suriano2018}.

\begin{acknowledgements}

The authors gratefully acknowledge the support of the Max Planck Society and the Swiss National Science Foundation (SNSF) through grant number 234038. We sincerely thank the technicians at MPIK, especially Benjamin Gramlich, Dennis Gro\ss{}, Michael Rei\ss{}felder and Jonas Westermann. We are indebted to Leander Fischer, Maja Lecher, Margherita Noia and Mona Piotter and for their contributions as part of their Bachelor’s and Master’s theses. We would also like to thank Dr. Mahdi Mahajeri (EuropCoating), Dr. Stefan Laure (Dr. Laure Plasmatechnologie GmbH), and Dr. St\'ephane Lucas (Innovative Coating Solutions) for their contributions, technical discussions, and assistance throughout the project. Furthermore, we gratefully acknowledge Dr. Tarek Lutz and his team at NMI Reutlingen for providing insight into our coating layers via FIB/SEM imaging.

\end{acknowledgements}

\bibliographystyle{utphys}
\bibliography{manuscript_v0.bib}   

\end{document}